\documentclass{jkasarchive}

\def\beginpage{1} % first page of article
\def\received{February 30, 2014} % date paper was received by JKAS
\def\accepted{February 31, 2014} % date of acceptance
\date{Received \received; accepted \accepted}

\usepackage{epstopdf}
\usepackage{amsmath}
\usepackage{rotating}
\title{
Intra-night Optical Variability of Active Galactic Nuclei in the COSMOS Field with the KMTNet
}

\author[1]{Joonho Kim}
\author[1,2]{Marios Karouzos}
\author[1]{Myungshin Im}
\author[1]{Changsu Choi}
\author[1]{Dohyeong Kim}
\author[3]{Hyunsung D. Jun}
\author[4,5]{Joon Hyeop Lee}
\author[6,7]{Mar Mezcua}

\affil[1]{Center for the Exploration of the Origin of the Universe (CEOU), Astronomy Program, Department of Physics and Astronomy, Seoul National University, Seoul 08826, Korea;}
\affil[ ]{\email{joonho@astro.snu.ac.kr, mim@astro.snu.ac.kr}}
\affil[2]{Nature Astronomy, Springer Nature, 4 Crinan Street, N1 9XW London, United Kingdom}
\affil[3]{School of Physics, Korea Institute for Advanced Study, 85 Hoegiro, Dongdaemun-gu, Seoul 02455, Korea}
\affil[4]{Korea Astronomy and Space Science Institute, Daejeon 34055, Korea}
\affil[5]{University of Science and Technology, Daejeon 34113, Korea}
\affil[6]{Institute of Space Sciences (ICE, CSIC), Campus UAB, Carrer de Magrans s/n, 08193 Barcelona, Spain}
\affil[7]{Institut d'Estudis Espacials de Catalunya (IEEC), C/ Gran Capit\`{a}, 08034 Barcelona, Spain}
\def\corrauthor{
Myungshin Im
}

\def\runningauthor{
J. Kim et al.
}

\def\runningtitle{
Intra-night optical variability of AGN in the COSMOS field with the KMTNet
}

\def\keywords{
KMTNet: AGN: optical: variability: microvariability: intranight
}

\def\abstracttext{
Active Galactic Nucleus (AGN) variability can be used to study the physics of the region in the vicinity of the central black hole. In this paper, we investigated intra-night optical variability of AGN in the COSMOS field in order to understand the AGN instability at the smallest scale. Observations were performed using the KMTNet on three separate nights for 2.5-5 hour at a cadence of 20-30 min. We find that the observation enables the detection of the short-term variability as small as $\sim 0.02$ and 0.1 mag for $R \sim 18$ and 20 mag sources, respectively. Using four selection methods (X-rays, mid-infrared, radio, and matching with SDSS quasars), 394 AGNs are detected in the 4 deg$^2$ field of view. After differential photometry and $\chi^2-$test, we classify intra-night variable AGNs. But the fraction of variable AGNs (0-8 \%) is consistent with a statistical fluctuation from null result. Eight out of 394 AGNs are found to be intra-night variable in two filters or two nights with a variability level of 0.1 mag, suggesting that they are strong candidates for intra-night variable AGNs. Still they represent a small population (2 \%). There is no sub-category of AGNs that shows a statistically significant intra-night variability. 
}

\begin{document}
\jkashead %% set title, authors, abstract, etc.

%%%%%%%%%%%%%%%%%%%%%%%%%%%%%%%%%%%%%%%%%%%%%%%%%%%%%%%%%%%%%%%%%%%%%
%%% BEGIN MAIN TEXT HERE %%%%%%%%%%%%%%%%%%%%%%%%%%%%%%%%%%%%%%%%%%%%
%%%%%%%%%%%%%%%%%%%%%%%%%%%%%%%%%%%%%%%%%%%%%%%%%%%%%%%%%%%%%%%%%%%%%

\section{Introduction\label{sec1}}

The flux variability is a representative characteristic of active galactic nucleus (AGN). Because it is hard to resolve AGN, variability is widely used to study the size, mass, and structure of AGN. Observations have discovered that AGN variability appears not only in various wavelengths, but also shows various timescales from hours to years. The results show increase in variability amplitude at shorter wavelengths and longer timescales (\citealt{Ulrich1997} and references therein). In addition, there is an inverse correlation between long-term variability and luminosity while there is a positive correlation between long-term variability and redshift, as discovered by \citet{Uomoto1976} and \citet{Pica1983}.

Unfortunately, the origin of variability is not completely known. Generally, instability in the accretion disk or radio jet have been suggested as explanations for the AGN variability. For the accretion disk, physical timescales could be correlated with the variability of AGN. Hour-scale, day-scale, and year-scale variability correspond to light crossing timescale, dynamical timescale, and thermal timescale of the accretion disk, respectively (\citealt{Frank2002}). Therefore, the intra-night optical variability (INOV), which corresponds to the light crossing timescale, can be used to study violent variation at the smallest AGN size and structure. However, hour-scale variability has not been studied as much as longer timescale variability as it needs intensive monitoring with high cadence. It is also hard to detect, because of the minuscule variability from $\sim10^{-1}-10^{-3}$ mag.

Previous results on this subject present mixed results that require careful examination. This is because different types of AGNs show different INOV activities and the depths of the data vary between different studies. Several consensuses and controversial issues can be summarized as below.

First, sub-classes of AGNs such as radio-loud AGNs and blazars are found to have a high probability of exhibiting INOV. The ballpark number is 20-100\% for the fraction or the duty cycle of radio-loud quasars and blazars with INOV at the level of a few percent or more (\citealt{Jang1995,Jang1997}; \citealt{Romero1999,Romero2002}; \citealt{Stalin2005}; \citealt{Goyal2010,Goyal2013a}). On the other hand, radio-quiet AGNs show INOV much less frequently, but the results are more controversial than the case for radio-loud AGNs. In general, it is accepted that the INOV is found in none or only a few \% of radio-quiet AGNs of various kinds when the variability amplitude of a few percent is examined (\citealt{Romero1999}; \citealt{Webb2000}; \citealt{Klimek2004}; \citealt{Bachev2005}; \citealt{Carini2007}; \citealt{Kumar2016}), where the variability amplitude is defined as the ratio of the standard deviation of the AGN light curve versus the observational error being $> 2.576$ or so (99\% confidence; \citealt{Jang1997}). \citet{Webb2000} examined INOV of 20 Seyfert 1 AGNs, and found no INOV with the upper limit on the amplitude of the variation of 0.03 mag for time scales of 1 hr or less. However, they found the AGN variation at 25 to 27 hr level for two of their sample. Their photometric error is 0.02 mag. No INOV is found for 18 high luminosity quasars at $z > 2$ in which the photometric errors are of order of 0.02 to 0.05 mag (\citealt{Bachev2005}). At a similar photometric error accuracy, \citet{Klimek2004} found only one occasion of INOV of 0.045 mag change among 33 nights of the data for six Seyfert 1 galaxies, suggesting the duty cycle of 3\% at the level of 0.05 mag variability. Using photometric data at an accuracy of 0.02 mag, \citet{Carini2007} find no INOV among 7 quasars (5 radio-quiet and 2 radio-intermediate). \citet{Kumar2016} examined the INOV of 10 weak emission line quasars (WLQ) in the hope of identifying BL Lacs with a peculiar property, but they found no INOV among 18 observing sessions. Their work has photometric error of 0.03 to 0.06 mag, so INOV of 0.1 mag or larger should have been detected.

When smaller amplitudes of INOV are examined, the results are rather mixed. Examining the INOV amplitude down to about 1\%, several studies find 10-40\% of radio-quiet quasars show INOV with the duty cycle of 10-20\% (\citealt{Jang1995,Jang1997}; \citealt{Romero1999,Romero2002}; \citealt{Gopal-Krishna2000, Gopal-Krishna2003}; \citealt{Webb2000}; \citealt{Stalin2004a,Stalin2004b,Stalin2005}; \citealt{Klimek2004}; \citealt{Bachev2005}; \citealt{Gupta2005}; \citealt{Carini2007}; \citealt{Goyal2010,Goyal2012,Goyal2013a}; \citealt{Joshi2011}; \citealt{Kumar2016}), while several other studies that examined INOV at a similar level do not find such evidence (\citealt{Petrucci1999}; \citealt{Romero1999}). For example, \citet{Goyal2013a} find that about 40\% of radio-quiet quasars exhibit INOV, and 20\% if limited to the variability amplitude of $\sim$ 3\%. The duty cycle, on the other hand, is found to be 10\% (or 6\% at the variability amplitude $> 3$\%) for these objects. However, at a similar photometric accuracy, \citet{Romero1999} examined the INOV of 8 RQQs and RLQs, BL Lacs, finding no convincing case for INOV among radio-quiet quasars. At a slightly worse photometric accuracy (0.5\%), \citet{Petrucci1999} found no INOV among 22 Seyfert 1 galaxies.

More recently, the high cadence photometry data from the $Kepler$ mission (30 min cadence) are being studied to examine the short-term variability of AGNs (e.g, \citealt{Mushotzky2011}, \citealt{Carini2012}, \citealt{Wehrle2013}, \citealt{Revalski2014}, \citealt{Chen2015}, \citealt{Edelson2014}, \citealt{Kasliwal2015}, \citealt{Shaya2015}, \citealt{Dobrotka2017}). \citet{Aranzana2018} examined the variability of 252 AGNs with $R < 19$ mag. The power spectral density (PSD) of these objects in this study are consistent with white noise of a few tenths of a percent error for most objects, although a few objects seem to show $\sigma >$ 1 \% with the PSD following a power-law. However, no statistical inferences have been made about the fraction of AGNs for INOV.

Obviously our understanding of the demography of the INOV among AGNs can be improved considering the small sample sizes in most of the previous studies. Furthermore, inhomogeneously sampled data, observed at different nights, bands, and telescopes, have made it difficult to compare various results about the variability of AGN. The Korea Microlensing Telescope Network (KMTNet) is a network of three 1.6m telescopes that are designed to study microlensing events without a gap in observing times (\citealt{Kim2016}). This facility offers an excellent opportunity to study INOV of a large number AGNs of different kinds, and to do so, we are conducting a project named KMTNet Active Galactic Nuclei VAriability Survey (KANVAS). In this paper, we present the results from our pilot study of KANVAS. In Section \ref{sec2}, we introduce the KMTNet and observations while Section \ref{sec3} presents an analysis of KMTNet data. Section \ref{sec4} describes the AGN selection method, and their characteristics. Sections \ref{sec5} and \ref{sec6} present the variable AGN and their results. Finally, we discuss and conclude this results in Section \ref{sec7}.

%%% TABLE %%%%%%%%%%%%%%%%%%%%%%%%%%%%%%%%%%%%%%%%%%%%%%%%%%%%%%%%%%%%%%%%%%%%%
\begin{table*}[t!]
\centering
\caption{KMTNet-CTIO observations on the COSMOS field.}
\begin{tabular}{lcccccc}
\toprule
Date & Band & Duration time & Total exposure time & \# Epoch & Seeing & 5$\sigma$ Depth [mag]\\
yyyy.mm.dd & & & per epoch & & & per epoch\\
\midrule
2015.02.18 & $B$ & $\sim$2.5h & 360s & 5 & $1.5 - 2.0''$ & 22.5 - 22.7 \\
 & $V$ & $\sim$2.5h & 240s & 5 & $1.5 - 1.8''$ & 22.2 - 22.4 \\
 & $R$ & $\sim$2.5h & 180s & 5 & $1.4 - 1.8''$ & 21.8 - 22.3 \\
 & $I$ & $\sim$2.5h & 180s & 5 & $1.4 - 1.9''$ & 21.1 - 21.4 \\
2015.03.21 & $B$ & $\sim$5h & 600s & 13 & $2.0 - 2.3''$ & 21.4 - 22.6 \\
 & $R$ & $\sim$5h & 270s & 13 & $1.5 - 2.1''$ & 20.4 - 22.1 \\
2016.04.08 & $B$ & $\sim$4h & 600s & 11 & $1.6 - 2.1''$ & 22.7 - 23.0 \\
 & $R$ & $\sim$4h & 270s & 11 & $1.3 - 1.9''$ & 21.9 - 22.3 \\
\bottomrule
\label{tab1}
\end{tabular}
\end{table*}
%%%%%%%%%%%%%%%%%%%%%%%%%%%%%%%%%%%%%%%%%%%%%%%%%%%%%%%%%%%%%%%%%%%%%%%%%%%%%%%

\section{Observation\label{sec2}}

The main research program of the KMTNet is to monitor galactic bulge and discover exoplanets. However, in seasons when the galactic bulge cannot be observed, other science programs were allocated observing time. Our study is one of these non-bulge science programs of the KMTNet. The KMTNet is composed of three 1.6m telescopes located in Chile (CTIO), South Africa (SAAO), and Australia (SSO), with a 4 deg$^2$ field of view and $0.4''$ pixel size. The 24 hour, wide-field monitoring system allows all-day continuous monitoring of AGN.

In this paper we present observations of the COSMOS field (10:00:28.6 +02:12:21) using the KMTNet. For this pilot study, we use three nights worth of data that were taken at the CTIO station. The first observation was 2015 February 18 where we obtained the COSMOS field data for $\sim$2.5 hours of observation using $B$, $V$, $R$, and $I$ bands. On 2015 March 21, we observed the COSMOS field again for $\sim$5 hours using $B$ and $R$ bands. Finally, on 2016 April 8, another observation was performed for $\sim$4 hours in $B$ and $R$ bands. Two to five consecutive frames, each with 1.5-2 min exposure time were taken in each band sequentially. The 2-5 frames are combined to produce a deeper single epoch image with a total exposure time of 3-10 min. Approximately, 5-13 sets of these data were taken, to produce 20-30 min cadence data per filter. The seeing conditions, which were measured using SExtractor,
were $\sim1.6''$ on 2015 February 18, $\sim2.0''$ on 2015 March 21, and $\sim1.8''$ on 2016 April 8. However, the weather condition was very unstable on 2015 March 21 and we excluded 4 epochs of data. Furthermore, the depth of the March data is shallower than for other dates and the observation log is provided in Table \ref{tab1}.

\section{Data analysis\label{sec3}}

The basic data reduction, such as bias correction and flat-fielding, was done by the KMTNet pipeline at the KMTNet data center. We then performed astrometry, image stacking, and photometry calibration on the reduced data. First, we conducted astrometry following the basic process provided by the KMTNet team - Astrometric Calibration for the KMTNet Data\footnote{\url{http://kmtnet.kasi.re.kr/kmtnet-eng/astrometric-calibration-for-kmtnet-data}}. We used SExtractor (\citealt{Bertin1996}), SCAMP (\citealt{Bertin2006}), and SWarp (\citealt{Bertin2002}) software packages for astrometry and achieved astrometric accuracy of $\sim0.2''$. SWarp is used again for co-addition of 2-5 frames that were taken consecutively at a given epoch. Because seeing condition varied during the night, systematic bias could be introduced when fixed aperture is used for photometry. To make images have similar seeing conditions, we applied Gaussian convolution based on the seeing FWHM of each image (i.e., downgrade the seeing FWHM of all images to the worst seeing image). Then, SExtractor was used again with S/N = 5 limit to produce source catalogs of the convolved images. In the 2015 February observation, there were $\sim$50,000, $\sim$65,000, $\sim$90,000, and $\sim$80,000 detected sources in $B$, $V$, $R$, and $I$ bands, respectively, with a median photometric uncertainty of $\sim$0.1 mag. For the photometry calibration of the detected sources, we used the magnitudes of stellar sources from the photometric catalog of the COSMOS field (\citealt{Capak2007}). For the 2015 February data, the image depths at 5$\sigma$ detection (point source) are 22.6, 22.3, 22.2, and 21.3 magnitudes in $B$, $V$, $R$, and $I$ bands, respectively. Hereafter, we only use $8''$ aperture magnitude. This aperture covers $>95$\% of total flux and it corresponds to 3-4 times of the maximum seeing FWHM.

\section{AGN Samples\label{sec4}}

We selected AGNs in X-ray, mid-infrared, radio, and optical bands. The multi-wavelength catalogs and our source catalogs are matched using 1 arcsec radius and the closest source within the radius is selected. Typically, between different catalogs, the coordinate differences are at the level of 0.2 to 0.4 arcsec for the matched sources. AGNs are excluded from our selection in cases of saturation or column pattern in the KMTNet image, proximity to a bright source, or if any other source is within the aperture size used for photometry. In addition, we only selected AGNs where the photometric error is smaller than 0.2 mag in our data. The following subsections describe AGN selection at different wavelengths. Since the number of selected AGNs depends on the different depths for each observed night and band, we provide ranges for the number of detected AGNs.

%%% TABLE %%%%%%%%%%%%%%%%%%%%%%%%%%%%%%%%%%%%%%%%%%%%%%%%%%%%%%%%%%%%%%%%%%%%%
\begin{table}[t]
\centering
\caption{Classification of X-ray selected AGNs in $R$ band of the 2016 April data.}
\begin{tabular}{lrrrrrrrrrrrrrrr}
\toprule
Class & Unobsc & Obsc & no Class & Total \\
\midrule
Broad-line AGN & 55 & 1 & 0 & 56 \\
non Broad-line AGN & 1 & 4 & 0 & 5 \\
no Class & 3 & 1 & 3 & 7 \\
Total & 59 & 6 & 3 & 68 \\
\bottomrule
\end{tabular}
\tabnote{obsc = obscured AGN}
\label{tab2}
\end{table}
%%%%%%%%%%%%%%%%%%%%%%%%%%%%%%%%%%%%%%%%%%%%%%%%%%%%%%%%%%%%%%%%%%%%%%%%%%%%%%%

\subsection{X-ray selected AGNs}

X-ray selected AGNs are taken from \citet{Civano2012}, where they classified AGNs as broad line AGN (BLAGN) or not (non-BLAGN) if spectra were available, and obscured or unobscured AGN based on fitting their spectral energy distribution. There are 549 AGNs identified in the 0.9 deg$^2$ field of view covered by the Chandra-COSMOS catalog, but due to the depth of the KMTNet observations, only about 25-90 X-ray AGNs are detected in the KMTNet images, depending on the date of the observing run and the filter. Table \ref{tab2} shows an example of the X-ray detected AGN classification for the 2016 April data (also see Table \ref{tab3} for the number of X-ray matched AGNs). Roughly $\sim80$\% of them are unobscured broad-line AGNs. The X-ray flux range in 0.5-10 keV is $2.6\times10^{-15} - 1.6\times10^{-13} ~ erg ~ cm^{-2} ~ s^{-1}$ while the X-ray luminosity is $5.4\times10^{41} - 2.0\times10^{44} ~ erg ~ s^{-1}$.

\subsection{Mid-Infrared selected AGNs}

\begin{figure}[t]
\centering
\includegraphics[width=90mm,angle=270]{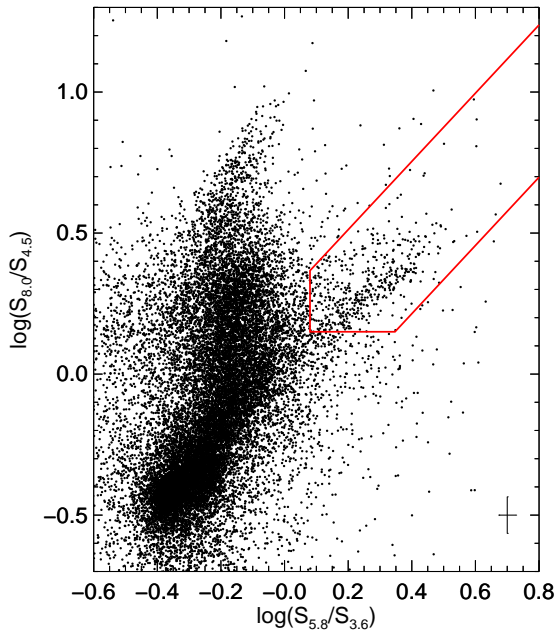}
\caption{Mid infrared color-color diagram of the KMTNet detected sources in the COSMOS field. The x-axis is the ratio of 5.8 $\mu$m and 3.6 $\mu$m fluxes, while the y-axis shows the ratio of the 8.0 $\mu$m and 4.5 $\mu$m fluxes. Sources within the red solid lines are selected as AGNs \citealt{Donley2012}. The median error of the points is shown in the bottom-right corner.}
\label{fig1}
\end{figure}

\citet{Lacy2004} and \citet{Stern2005} suggest mid-infrared (MIR) color-color selection of AGN with Spitzer IRAC photometry. This method can effectively select more obscured AGN, though deeper IRAC data reveal contamination from normal galaxies. Here, we employed the recent selection method defined by \citet{Donley2012} to minimize contamination. From the Spitzer-COSMOS IRAC 4-channel photometry catalog (\citealt{Sanders2007}) which covers 2 deg$^2$ of the COSMOS field, aperture corrected $1.9''$ photometry is used at the 5$\sigma$ sensitivity limit in all channels. Consequently, 64-174 MIR selected AGNs are detected in our data, depending on the date of the observation and the filter (Figure \ref{fig1}).

\subsection{Radio selected AGNs}

\begin{figure}
\centering
\includegraphics[width=58mm,angle=270]{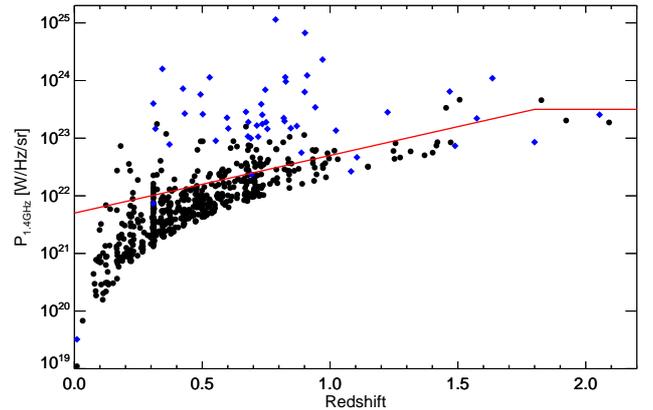}
\caption{The distribution of 1.4GHz radio power (x-axis) and redshift (y-axis) of the KMTNet detected sources in the COSMOS field. Radio-loud sources are shown with blue diamonds, while radio-quiet sources are indicated with black circles. The red solid line is the radio power cut used for the radio selection of AGNs. Sources above the red solid line are selected as radio AGNs. The radio power uncertainties are smaller than symbol size.}
\label{fig2}
\end{figure}

Radio sources in general are either AGNs or star-forming galaxies, but bright radio sources are known to be mostly AGNs. At the same time, the brightness of radio sources is known to evolve as a function of redshift. To differentiate AGNs from star-forming galaxies, \citet{Magliocchetti2014} suggest a selection based on the radio power in Equation (\ref{eq1}),
\begin{equation}
\begin{split}
& P_{1.4GHz} > 10^{21.7+z} ~ [W/Hz/sr] ~~~~~ if ~~ z\le1.8\\
& P_{1.4GHz} > 10^{23.5} ~~~~ [W/Hz/sr] ~~~~~ if ~~ z>1.8 \\
\end{split}
\label{eq1}
\end{equation}
Here, the radio power at 1.4 GHz, $P_{\rm{1.4 GHz}}$, is calculated as 
\begin{equation}
\begin{split}
P_{\rm{1.4 GHz}} = F_{\rm{1.4 GHz}} D_{\rm{L}}^{2} \times (1+z)^{\alpha-1},
\end{split}
\label{eq2}
\end{equation}
where $F_{\rm{1.4 GHz}}$ is the observed flux at 1.4 GHz, $D_{\rm{L}}$ is the luminosity distance, and $\alpha$ is the power law index of the radio spectrum $F_{\nu} \sim \nu^{-\alpha}$, and a canonical value of $\alpha=0.7$ is chosen here (e.g., \citealt{Smolcic2014,Smolcic2017}). This $(1+z)$ term is included for the K-correction. Radio AGNs in the COSMOS field are selected by applying this method to the 1.4 GHz radio flux in the VLA-COSMOS catalog (\citealt{Schinnerer2007}) covering 2 deg$^2$ field of view. To calculate the radio power, we used the spectroscopic redshift catalog (\citealt{Lilly2009}, \citealt{Damjanov2018}) and photometric redshift catalog (\citealt{Ilbert2009}) of the COSMOS field by assuming isotropic radio emission. This selection yields 14-61 AGNs, depending on the date of the observation and the filter. Figure \ref{fig2} shows the radio power versus redshift of radio sources matched with the KMTNet detected sources, as well as the radio selection criteria defined by Equation (\ref{eq1}). Furthermore, we calculated the radio loudness using Equation (\ref{eq3}), where we classify radio-loud (quiet) sources as those where the radio loudness is larger (smaller) than 1 (\citealt{Ivezic2002}). The fraction of radio-loud sources is 30-60\% in our radio AGN sample.
\begin{equation}
\begin{split}
R_{i} = \rm{log} (F_{\rm{1.4 GHz}}/F_{\rm{i-band}})
\end{split}
\label{eq3}
\end{equation}
\begin{figure}[t!]
\centering
\includegraphics[width=75mm,angle=270]{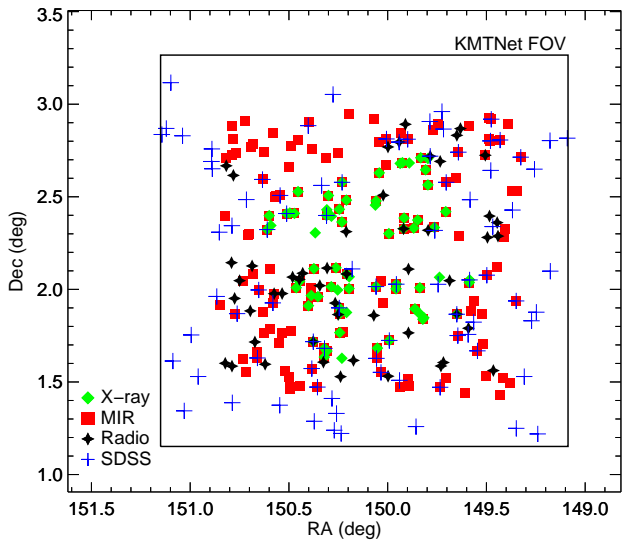}
\caption{Positions of the selected AGNs from the 2016 April data in $R$ band. Green diamonds, red squares, black stars, and blue crosses correspond to the X-ray selected AGNs, the MIR selected AGNs, the radio selected AGNs, and the SDSS quasars, respectively.}
\label{fig3}
\end{figure}
\begin{figure}[t!]
\centering
\includegraphics[width=75mm,angle=270]{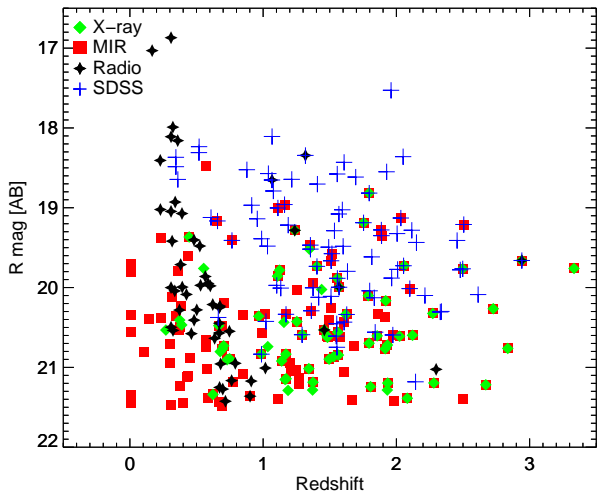}
\caption{The apparent magnitude in $R$ band and redshift of selected AGNs from the 2016 April data. Green diamonds, red squares, black stars, and blue crosses correspond to X-ray selected AGNs, MIR selected AGNs, radio selected AGNs, and matched SDSS quasars, respectively. The errors are smaller than the sizes of symbols.}
\label{fig4}
\end{figure}
\begin{figure}[t!]
\centering
\includegraphics[width=130mm,angle=270]{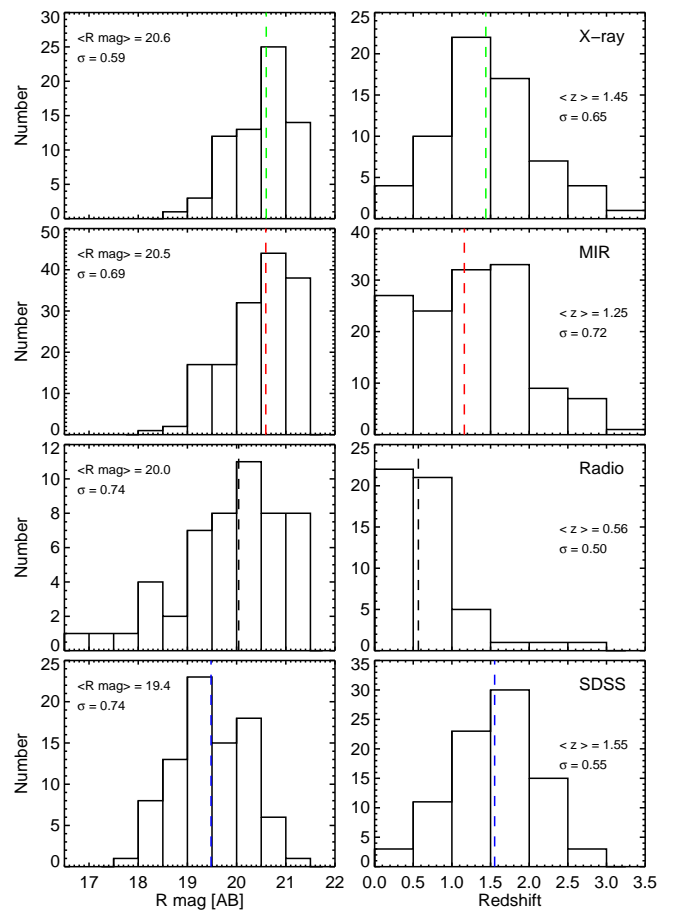}
\caption{The apparent magnitude (left) and redshift (right) histogram of AGNs from the 2016 April data in $R$ band. The median value (the vertical dashed line) and the dispersion of the distribution is given in each panel. Also given are the AGN selection methods in each right panel.}
\label{fig5}
\end{figure}

\subsection{SDSS DR7 quasars catalog}

Type 1 AGNs can be selected from optical colors and we use those AGNs that are identified in the SDSS DR7 quasar catalog (\citealt{Schneider2010}). They are luminous ($M_i < -22$), broad-line AGNs for which physical parameters such as M$_{BH}$ are available in the catalog. Furthermore, the quasar selection is not limited to the COSMOS field, and the SDSS type 1 quasars are also identified in the KMTNet field of view outside the COSMOS field. We identify 65-87 SDSS DR7 quasars in the KMTNet field of view, depending on the date of observation and the filter.

\subsection{Short summary on AGNs selected from different methods}

Figure \ref{fig3} shows the positions of selected AGNs from the 2016 April data in $R$ band. As shown in the figure, some AGNs are selected by more than one selection method. 68, 151, 51, and 85 AGNs are selected from the X-ray, Mid-infrared, Radio, and SDSS quasar catalogs, respectively, in this particular case. Figure \ref{fig4} shows the magnitude in $R$ band versus the redshift of the selected AGNs while Figure \ref{fig5} shows the distribution of apparent magnitude (left) and redshift (right). The apparent magnitudes of radio selected AGNs and SDSS quasars are brighter than MIR selected AGNs and X-ray selected AGNs. The median values of apparent magnitude are 20.6, 20.5, 20.0, and 19.4 for X-ray, MIR, Radio, and SDSS, respectively. In the case of redshift, X-ray selected AGNs and SDSS quasars have higher redshift than MIR selected AGNs and radio selected AGNs. The median values are 1.45, 1.25, 0.56, and 1.55 for X-ray, MIR, radio, and SDSS, respectively. The list of detected AGN is given in Table \ref{tab5} of Appendix A, where we list 394 AGNs that are detected in the KMTNet data at least once in one of the three observing runs and one filter.

\begin{figure}[t!]
\centering
\includegraphics[width=75mm,angle=270]{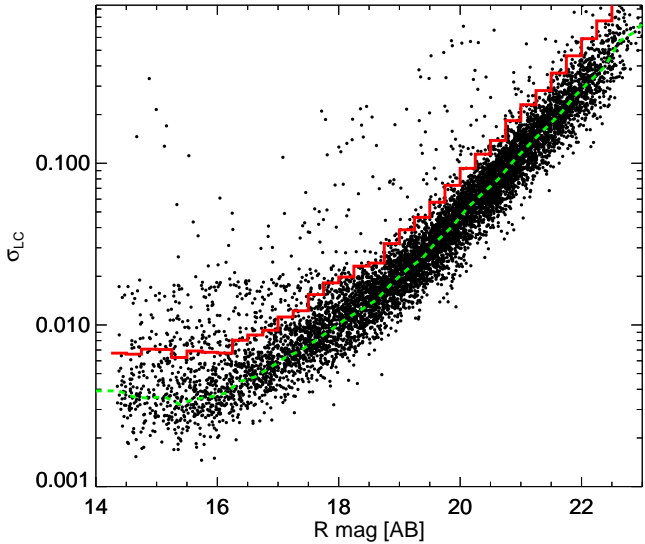}
\caption{The standard deviation of the multi-epoch $R$ band magnitudes for each star versus the mean $R$ band magnitude of each star over the observed period. A red solid line is 3-sigma cut of ${\sigma}_{LC}$ of stars in each 0.25 magnitude bin while a green solid line is median value of ${\sigma}_{LC}$. Stars below this red solid line are selected as non-variable stars.}
\label{fig6}
\end{figure}
\begin{figure}[t!]
\centering
\includegraphics[width=75mm,angle=270]{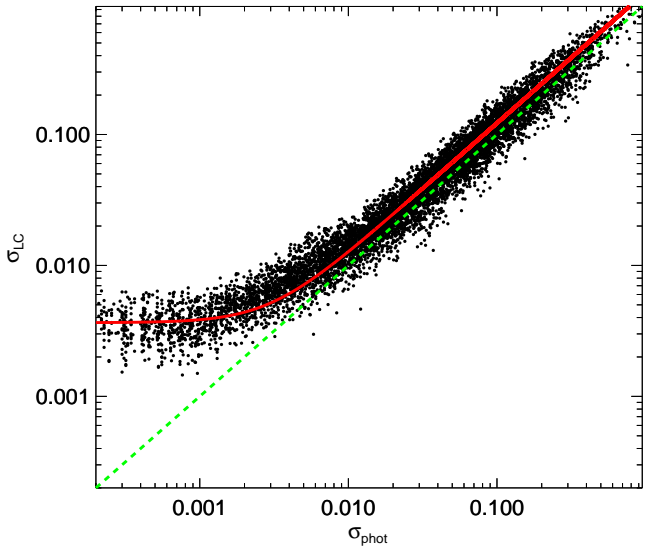}
\caption{The standard deviation of the multi-epoch $R$ band magnitudes for each star versus the mean photometric error of each star over the observed period. A green dashed line is the one-to-one line, and a red solid line is fitting result using Equation (\ref{eq4}). It shows underestimation of photometric error.}
\label{fig7}
\end{figure}

\section{Variability of AGN\label{sec5}}

%Star selection
Absolute photometry calibration is dependent on the photometry calibration error of the COSMOS catalog and can be uncertain by more than $\sim10^{-2}$ mag. For this reason, absolute calibration is attempted on images from one of the epochs with the best quality. The photometry from the absolute calibration is kept as a reference to judge the amount of variations in magnitude for data on other nights. To obtain the light curve with higher accuracy, we performed differential photometry for the selected AGNs. First, we identified stars within the COSMOS field using parameters in SExtractor by selecting sources with CLASS\_STAR (stellarity) greater than 0.95, and FLAGS= 0. To select non-variable stars only, we performed differential photometry for each selected star using $\sim$20 other bright ($<$ 16 mag) stars within 0.2 degree. The magnitude standard deviations of differential light-curves (${\sigma}_{LC}$) for all stars are calculated. Then, we chose stars within 3-sigma cut of ${\sigma}_{LC}$ in each 0.25 magnitude bin as non-variable stars (Figure \ref{fig6}). A few thousands to ten thousand non-variable stars are selected in each band and each night.

%Error calibration
Figure \ref{fig6} shows an interesting trend where the $\sigma_{LC}$ does not decrease below $0.003$ mag even at very bright magnitudes ($<$ 16 mag). Seen in a different way, Figure \ref{fig7} shows $\sigma_{LC}$ versus $\sigma_{phot}$, the error from the photometry measurement. The trend seen in Figure \ref{fig6} persists, i.e., $\sigma_{LC}$ staying above 0.003 mag even when $\sigma_{phot}$ goes well below 0.003 mag. This trend suggests that there is a minimum photometric error possibly due to flat-fielding error, systematic uncertainties due to frame-to-frame photometry variation, geometric distortion effect, and other effects. An accurate estimation of photometric error is crucial when we classify variable AGN in order to minimize the number of false positives. Therefore, we set $\sigma_{int}$ as the minimum intrinsic error, below which we cannot reduce the photometry error in the KMTNet data. Another source of error is identified at large $\sigma_{phot}$ ($>0.01$ mag) where $\sigma_{int}$ is negligible. In this regime, $\sigma_{phot}$ is found to be smaller than $\sigma_{LC}$ by a factor of 1.2 to 1.4. We trace the cause of the difference to the re-sampling process in SWarp of sub-pixel shifted images for stacking that decreases background noise of the stacked image (\citealt{Bertin2002}). Note that a similar factor has been introduced in other AGN variability studies (e.g., \citealt{Goyal2013b}). Thus, we used Equation (\ref{eq4}) to model the true photometry error $\sigma_{cal}$, as
\begin{equation}
\begin{split}
& {\sigma}_{cal}=\sqrt{{{\sigma}_{int}}^{2}+(\eta\times{{\sigma}_{phot}}^{2})}
\end{split}
\label{eq4}
\end{equation}
\begin{figure}[t!]
\centering
\includegraphics[width=190mm,angle=270]{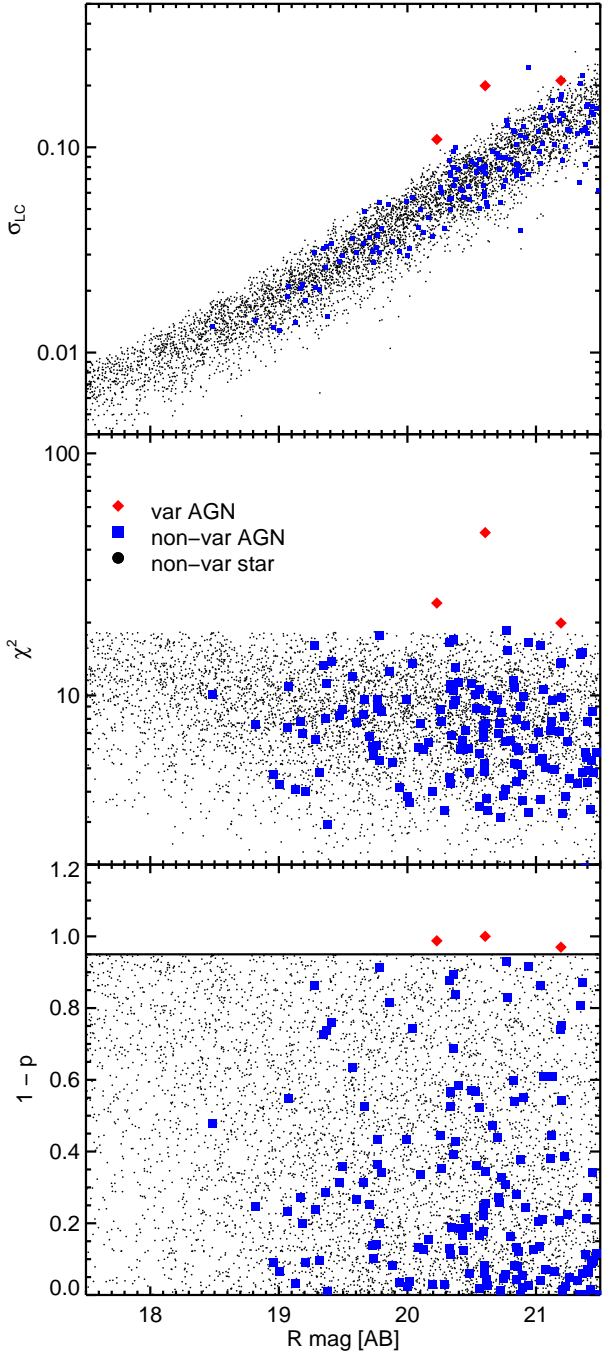}
\caption{${\sigma}_{LC}$ (top), $\chi^2$ (middle) and 1-p (bottom) with their apparent magnitude of MIR selected AGNs in $R$ band of the 2016 April data (dof = 10 for the 11 epochs data). Red diamonds, blue squares, and black circles correspond to variable AGN, non-variable AGN, and non-variable stars, respectively. The $\chi^2$ are calculated by Equation (\ref{eq5}). A black horizontal line denotes 1-p=0.95 and sources over this line are classified as variable AGNs.}
\label{fig8}
\end{figure}

\noindent where $\eta$ is an underestimation factor and ${\sigma}_{phot}$ is the mean photometric error of an object over all epochs. The intrinsic error and underestimation factor are estimated from each band and each night by fitting the $\sigma_{LC}$ versus $\sigma_{phot}$ relation as in Figure \ref{fig7}. The typical $\sigma_{int}$ value is 0.002-0.005 and the underestimation factor ranges between 1.2 and 1.4.

%Differential photometry
After the error calibration, we constructed the light curves of AGNs by differential photometry using the closest (within 0.2 degree) $\sim20$ non-variable reference stars brighter than 16 mag for each corresponding filter. To detect variability, we used the $\chi^2-$test (\citealt{deDiego2010}, \citealt{Villforth2010}). The $\chi^2$ value is calculated using Equation (\ref{eq5}).
\begin{equation}
\begin{split}
& {\chi}^{2} = \sum_{i} \frac{({m}_{i}-<{m}_{i}>)^{2}}{({\sigma}_{cal,i})^{2}}
\end{split}
\label{eq5}
\end{equation}

\noindent Here, $m_i$ denotes magnitude of each epoch, $<m_i>$ is average magnitude over all epochs, and ${\sigma}_{cal,i}$ is calibrated photometric error of each epoch. The p-value is calculated from the $\chi^2$-distribution. AGNs are classified as variable when $1-p > 0.95$, meaning that we reject the null hypothesis of non-variability at the 95\% confidence level ($\alpha=0.05$). Figure \ref{fig8} shows ${\sigma}_{LC}$, $\chi^2$ (degree of freedom, dof = 10) and $1-p$ of AGNs from the 2016 April data in $R$ band. We applied the $\chi^2-$test for comparison stars to examine the validity of the $\chi^2-$test. Except for the 2015 March data, the results show that 5-7\% of stars are classified as variable, which is expected according to our variability criterion. The 2015 March data shows $\sim$12\% fraction of variable stars, but we attribute this to the bad weather conditions during the 2015 March run. We also employ C-test (\citealt{Jang1997}, \citealt{Romero1999}) as a alternative statistical test. The 'C' statistic is defined as Equation (\ref{eq6}),
\begin{equation}
\begin{split}
& C = \frac{{\sigma}_{mag}}{<{\sigma}_{cal,i}>},
\end{split}
\label{eq6}
\end{equation}
where $<{\sigma}_{cal,i}>$ is the average photometric error over all the epochs. If the C value is greater than 1.95 (2.56), AGN can be classified as variable with 95\% (99\%) confidence level, assuming that the errors follow a Gaussian distribution. However, the $\chi^2$-test has an advantage that it uses photometric errors of each individual epoch unlike the C-test that uses the averaged photometric error. For that reason and since the C-test result is very similar to the $\chi^{2}$-test results, we mainly present the $\chi^{2}$-test results. The $1-p$ and $C$-statistic values of the whole sample of 394 AGNs are listed in Table \ref{tab5} of Appendix A.

%%% TABLE %%%%%%%%%%%%%%%%%%%%%%%%%%%%%%%%%%%%%%%%%%%%%%%%%%%%%%%%%%%%%%%%%%%%%
\begin{table}
\centering
\caption{The number of detected and variable AGNs with their fraction in the bracket.}
\begin{tabular}{lrrrrrrrrrrrrrrr}
\toprule
 & 2015.02.18 & 2015.03.21 & 2016.04.08 \\
\midrule
\multicolumn{4}{c} {X-ray selected AGN} \\
\midrule
$B$ & 2/78 (2.6\%) & 0/47 (0\%) & 1/80 (1.3\%) \\
$V$ & 2/85 (2.4\%) &  & \\
$R$ & 2/90 (2.2\%) & 0/25 (0\%) & 3/68 (4.4\%) \\
$I$ & 2/61 (3.3\%) &  & \\
\midrule
\multicolumn{4}{c} {MIR selected AGN} \\
\midrule
$B$ & 5/151 (3.3\%) & 1/91 (1.1\%) & 7/153 (4.6\%) \\
$V$ & 6/169 (3.6\%) &  & \\
$R$ & 5/174 (2.9\%) & 5/64 (7.8\%) & 3/151 (2.0\%) \\
$I$ & 6/117 (5.1\%) &  & \\
\midrule
\multicolumn{4}{c} {Radio selected AGN} \\
\midrule
$B$ & 1/32 (3.1\%) & 0/14 (0\%) & 2/27 (7.4\%) \\
$V$ & 3/47 (6.4\%) &  & \\
$R$ & 4/61 (6.6\%) & 0/24 (0\%) & 4/51 (7.8\%) \\
$I$ & 4/61 (6.6\%) &  & \\
\midrule
\multicolumn{4}{c} {SDSS DR7 quasars} \\
\midrule
$B$ & 4/80 (5.0\%) & 4/85 (4.7\%) & 7/87 (8.0\%) \\
$V$ & 2/83 (2.4\%) &  & \\
$R$ & 4/75 (5.3\%) & 4/65 (6.2\%) & 4/85 (4.7\%) \\
$I$ & 3/72 (4.2\%) &  & \\
\midrule
\bottomrule
\end{tabular}
\label{tab3}
\end{table}

\section{Results\label{sec6}}

Using the $\chi^2$-test, we could differentiate variable AGNs from non-variable AGNs. Table \ref{tab3} shows the number of detected and variable AGNs with their fraction in the bracket. As can be seen in Table \ref{tab3}, the fractions of variable AGNs do not significantly exceed the 5\% level that we found for the comparison stars. The results of the C-test at 95\% confidence level is very similar to the $\chi^2$-test results (not listed in Table \ref{tab3}). We can conclude that many of the AGN that are classified as intra-night variable are unlikely to be truly intra-night variable.

%%% TABLE %%%%%%%%%%%%%%%%%%%%%%%%%%%%%%%%%%%%%%%%%%%%%%%%%%%%%%%%%%%%%%%%%%%%%
\begin{table*}[t!]
\centering
\caption{Information of eight variable AGNs.}
\begin{tabular}{lrrrrrrrrrrrrrrr}
\toprule
Name & Selection & Redshift & Classification & Band & Mag & ${\sigma}_{var}$ (${\sigma}_{cal}$) & ${\sigma}_{var}$ (${\sigma}_{cal}$) & ${\sigma}_{var}$ (${\sigma}_{cal}$) \\
& & & and & & & 15.02.18 & 15.03.21 & 16.04.08 \\
& & & $\rm{log}(M_{\rm{BH}}/M_{\odot}))$ & & & 15.02.18 & 15.03.21 & 16.04.08 \\
\midrule
100008.9+021440 & X-ray & 2.536$^{a}$ & Broad-line & $B$ & 20.3 & \textbf{0.130} (0.039) & 3 & 2 \\
& \&MIR & & Obscured & $V$ & 20.0 & \textbf{0.172} (0.038) & & \\
& & & Radio-quiet & $R$ & 19.4 & \textbf{0.115} (0.027) & 3 & 2 \\
& & & & $I$ & 19.0 & 3 & & \\
\midrule
095820.7+020213 & X-ray & 1.856$^{a}$ & Broad-line & $B$ & 20.8 & -0.029 (0.077) & -0.100 (0.175) & \textbf{0.051} (0.054) \\
& \&MIR & & Unobscured & $V$ & 21.0 & -0.061 (0.115) & & \\
& & & & $R$ & 20.8 & 0.087 (0.142) & 1 & \textbf{0.172} (0.101) \\
& & & & $I$ & 21.0 & 1 & & \\
\midrule
100151.1+020032 & X-ray & 0.967$^{c}$ & Broad-line & $B$ & 20.7 & 3 & -0.117 (0.130) & -0.045 (0.057) \\
& \&MIR & & Unobscured & $V$ & 20.3 & \textbf{0.097} (0.057) & & \\
& & & & $R$ & 20.3 & \textbf{0.142} (0.064) & -0.169 (0.188) & -0.017 (0.068) \\
& & & & $I$ & 20.0 & 0.101 (0.122) & & \\
\midrule
095834.0+024427 & MIR  & 1.887$^{b}$ & Broad-line & $B$ & 19.5 & \textbf{0.025} (0.020) & -0.027 (0.047) & \textbf{0.034} (0.017) \\
& \&SDSS & & \& & $V$ & 19.4 & -0.019 (0.024) & & \\
& & & 8.88$^{d}$ & $R$ & 19.3 & 0.010 (0.025) & -0.049 (0.084) & 0.018 (0.025) \\
& & & & $I$ & 18.9 & 0.027 (0.044) & & \\
\midrule
100021.7+020000 & Radio & 0.219$^{c}$ & Radio-quiet & $B$ & 20.0 & \textbf{0.278} (0.042) & 2 & 2 \\
& & & \&& $V$ & 18.6 & \textbf{0.190} (0.014) & & \\
& & & Early-type & $R$ & 17.8 & 3 & 2 & 2 \\
& & & Galaxy & $I$ & 17.6 & 3 & & \\
\midrule
%100008.1+024554 & Radio & 0.029$^{c}$ & Radio quiet & B & 16.6 & 3 & \textbf{0.009} (0.005) & \textbf{0.008} (0.004) \\
%& & & \&& V & 16.1 & -0.002 (0.003) & & \\
%& & & Barred spiral & R & 15.7 & -0.002 (0.003) & \textbf{0.024} (0.005) & \textbf{0.003} (0.004) \\
%& & & Galaxy & I & 15.2 & -0.001 (0.003) & & \\
%\midrule
100017.7+030524 & SDSS & 2.178$^{b}$ & Broad-line & $B$ & 20.6 & \textbf{0.065} (0.046) & -0.029 (0.111) & 2 \\
& & & \& & $V$ & 20.8 & -0.019 (0.075) & & \\
& & & 8.38$^{d}$ & $R$ & 20.7 & \textbf{0.104} (0.087) & 1 & 2\\
& & & & $I$ & 20.1 & -0.050 (0.098) & & \\
\midrule
100434.3+025011 & SDSS & 1.215$^{b}$ & Broad-line & $B$ & 18.7 & 3 & 3 & \textbf{0.051} (0.011) \\
& & & \& & $V$ & 18.4 & \textbf{0.057} (0.011) & & \\
& & & 8.94$^{d}$ & $R$ & 18.4 & \textbf{0.063} (0.010) & 3 & \textbf{0.129} (0.017) \\
& & & & $I$ & 18.0 & 3 & & \\
\midrule
095621.6+024859 & SDSS & 0.956$^{b}$ & Broad-line & $B$ & 19.4 & 2 & 2 & \textbf{0.081} (0.017) \\
& & & \& & $V$ &  & 2 & & \\
& & & 8.99$^{d}$ & $R$ & 19.1 & 2 & 2 & \textbf{0.222} (0.024) \\
& & & & $I$ &  & 2 & & \\
\midrule
\bottomrule
\end{tabular}
\tabnote{1: Too faint, 2: Out of field of view, 3: Contaminated}
\tabnote{a: \citealt{Civano2012}, b: \citealt{Schneider2010}, c: \citealt{Abolfathi2017}, d: \citealt{Shen2011}}
\label{tab4}
\end{table*}
%%%%%%%%%%%%%%%%%%%%%%%%%%%%%%%%%%%%%%%%%%%%%%%%%%%%%%%%%%%%%%%%%%%%%%%%%%%%%%%

Nevertheless, the probability that an AGN is truly variable becomes higher if the AGN is classified as variable in more than two data sets. If one AGN is classified as variable in two bands or nights at the 95\% confidence level, then the expected probability of a false positive is only 0.25\%. We identify eight AGNs that are classified as intra-night variable in more than two bands or two nights both in the $\chi^2$-test and the C-test. We list these AGNs in Table \ref{tab4}. The AGN name comes from their coordinate information in the KMTNet catalog (hhmmss.s+ddmmss) and their magnitudes are the mean magnitude over all observed epochs. The variability strength is defined by the error subtracted magnitude variation as shown in Equation (\ref{eq7}), where ${{\sigma}_{mag}}$ denotes the standard deviation of the differential light-curve and ${{\sigma}_{cal}}$ is the mean calibrated photometric error. The variability strength of the variable AGNs based on the $\chi^2$-test is shown in bold. AGNs that were not selected for being too faint (a), out of field of view (b), and contamination by other sources or noise (c) are marked in the table.

\begin{equation}
\begin{split}
 {\sigma}_{var} = \sqrt{{{\sigma}_{mag}}^{2}-{{\sigma}_{cal}}^{2}} ~~~~~~~~~ if ~~{\sigma}_{mag} > {\sigma}_{cal} \\
 {\sigma}_{var} = -\sqrt{{{\sigma}_{cal}}^{2}-{{\sigma}_{mag}}^{2}} ~~~~~~~~ if ~~{\sigma}_{mag} < {\sigma}_{cal} \\
\end{split}
\label{eq7}
\end{equation}

\begin{figure*}[t!]
\centering
\includegraphics[width=217mm,angle=270]{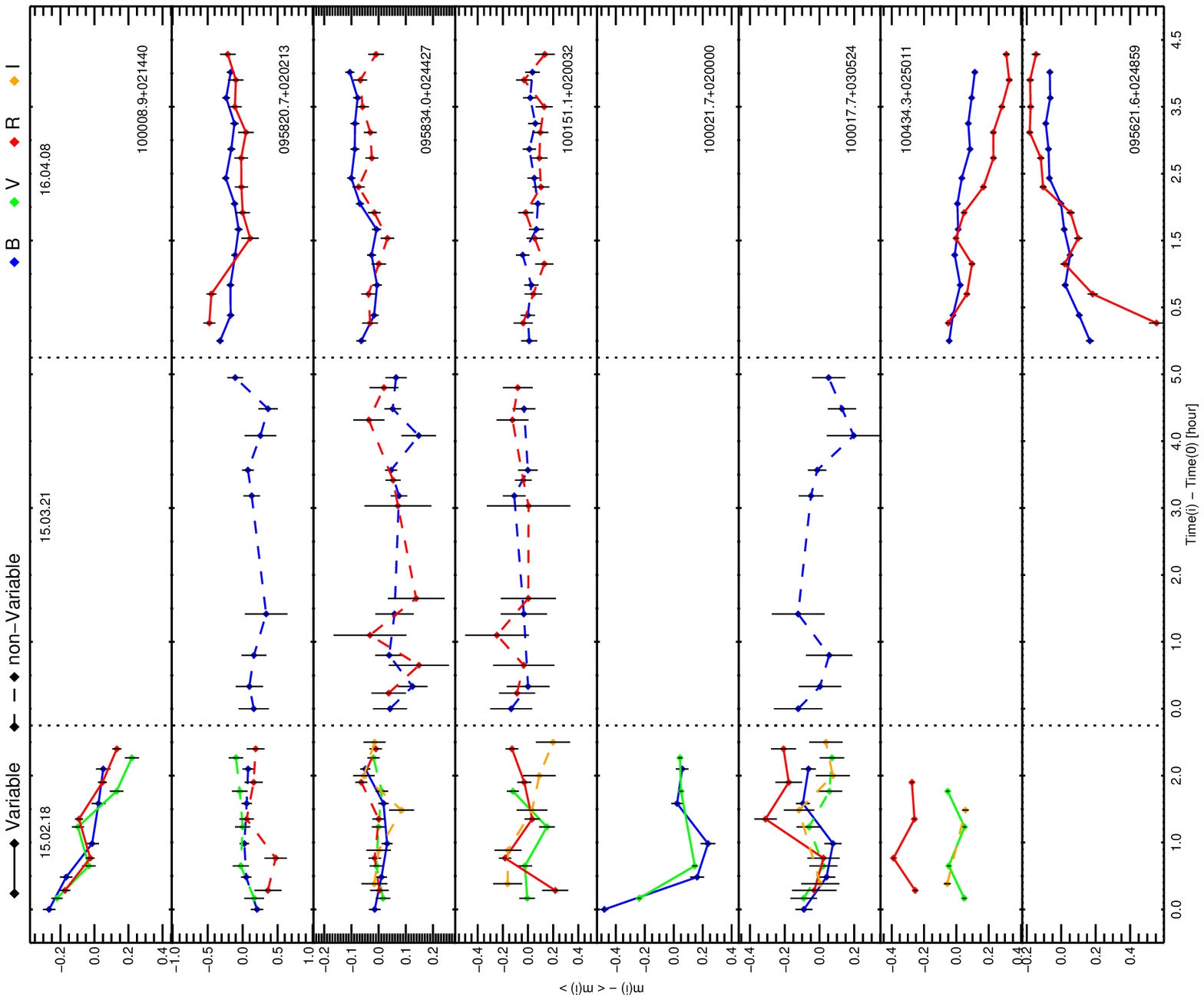}
\caption{The light-curves of eight variable AGNs. The name of AGN shown on the right side of each light-curve and the color denotes each band. Variable AGNs are plotted as solid lines while non-variable AGNs are plotted as dashed lines.}
\label{fig9}
\end{figure*}

In Figure \ref{fig9}, we plotted the light-curves of the eight variable AGNs. The AGN names are shown on the right side of the light-curve and the color denotes each band. Variable AGNs are plotted with solid lines while non-variable AGNs are plotted with dashed lines.\\

\textbf{100008.9+021440} This object is classified as a broad-line and obscured AGN in the Chandra-COSMOS identification catalog. The 2015 February data shows variability strength of 0.130 mag, 0.172 mag and 0.115 mag in $B$, $V$, and $R$ bands, respectively with 20.3 mag, 20.0 mag, and 19.4 mag in each band. Unfortunately, no usable data were obtained in 2015 March and 2016 April.

\textbf{095820.7+020213} This object is classified as a broad-line and unobscured AGN in the Chandra-COSMOS identification catalog. The variability of 0.05-0.17 mag is detected only in the 2016 April data, both in $B$ and $R$ bands.

\textbf{100151.1+020032} This is a X-ray and MIR-selected AGN which is classified as a broad-line and unobscured in the Chandra-COSMOS identification catalog. The variability is found only in the 2015 February data, with the $\sigma_{var} \sim 0.1-0.14$ in $V$ and $R$ bands.

\textbf{095834.0+024427} A small amount of variability, $\sim$0.03 mag, is detected in $B$ band in both the 2015 February and 2016 April data. This is a MIR-selected AGN which also appears in the SDSS quasar catalog.

\textbf{100021.7+020000} Radio-loud AGN with 20.0, and 18.6 magnitudes in $B$ and $V$ bands, respectively. This object is classified as variable in $B$ and $V$ bands with variability strength of 0.278 and 0.190 mag, respectively. However, the SDSS spectrum has the characteristics of early-type galaxy, not AGN. The SDSS image shows the morphology of an early-type galaxy too. The variability can arise from an increased nuclear activity or due to an unidentified systematic error in measuring the central variability of an extended source like this.

\textbf{100017.7+030524} SDSS quasar variable in $B$ and $R$ bands of the 2015 February data. The magnitudes are 20.6 and 20.7 in $B$, and $R$ bands with 0.065 and 0.104 mag variability strength, respectively.

\textbf{100434.3+025011} The brightest variable AGN in our sample. This SDSS quasar has apparent magnitudes of 18.7 and 18.4 in $B$ and $R$ bands, respectively. Variability is detected in both the 2015 February and 2016 April data with the strength of 0.06 to 0.13 magnitudes.

\textbf{095621.6+024859} This object is found to be variable in the 2016 April data, with the variability strengths of 0.08 and 0.22 and $B$ and $R$ bands, respectively.

\section{Discussion and Conclusion\label{sec7}}

In our pilot study, we identified eight intra-night variable AGNs out of 394 ($\sim2$\%). The smallness of the fraction can be due to the photometric error of the targets in our sample which is in the range 0.01 to 0.1 mag. In previous studies, \citet{Gupta2005}, \citet{Carini2007}, and \citet{Goyal2013a}, concluded that the INOV fraction of AGNs with 12-19 magnitudes is 10-30\%, but their photometric error is $\sim10^{-3}$ mag. In particular, \citet{Goyal2013a} show that peak-to-peak variabilities of most AGNs are less than 0.05 mag. We compared the peak-to-peak variability and the variability strength used in our study and we find the peak-to-peak variability corresponds to about three times the variability strength. The minimum variability strength of the eight INOV AGNs in our study is $\sim$0.03 mag (095834.0+024427), so we can detect a peak-to-peak variability of at least $\sim$0.09 mag. We checked with \citet{Goyal2013a} and find that the fraction of AGNs with the peak-to-peak variability greater than 0.09 mag is 4.8\%. This is roughly consistent with the 2\% in our results, considering that most of AGNs in this study have higher photometric uncertainty than 095834.0+024427. In addition, \citet{Webb2000}, \citet{Carini2003}, \citet{Klimek2004}, \citet{Bachev2005}, and \citet{Kumar2016}, show no variability or very small fraction of variability with $\sim10^{-2}$ mag photometric error. However, when investigating INOV of a very small amplitude (0.01 or less), we caution that there can be intrinsic errors such as the one we discussed in Section \ref{sec5}. In our case, the intrinsic error amounts to $0.004$ mag or so. As also discussed in \citet{Bachev2005}, ignoring this effect could lead to spurious detection of INOV. For now, we conclude that the low fraction of INOV AGNs indicates that the INOV amplitude is quite small if there is any -- smaller than our photometric error.

Several possible INOV mechanisms have been suggested: (i) the accretion disk instability; (ii) the X-ray irradiation of an accretion disk; (iii) the variation in jet activity such as a weak blazar component (See \citealt{Czerny2008}). AGNs with large $M_{\rm{BH}}$ are expected to show very weak INOV variability from the accretion disk instability (e.g., \citet{Bachev2005}; \citealt{Czerny2008}), but some INOV AGNs in our sample are found to have $M_{\rm{BH}}$ as massive as $10^{8}-10^{9}$ $M_{\odot}$. It will be interesting to see if there is a correlation between $M_{\rm{BH}}$ and the INOV variability as a way to support the accretion disk instability model. Otherwise, the accretion disk instabilities do not easily produce 0.1 mag level INOV observed in the 8 AGNs. For the X-ray irradiation instability, the variability time-scale is predicted to be a week or longer for high $M_{\rm{BH}}$ ($>$ $10^{8}$ $M_{\odot}$) AGNs. Furthermore, strong irradiation is required, which is not compatible with general broad-band spectral shape of AGNs (\citealt{Czerny2008}). Therefore, like the accretion disk instability model, the X-ray irradiation model faces a difficulty. This leaves the jet-related variability such as a weak blazar component as a plausible mechanism for the INOV of the 8 AGNs. The small fraction of INOV AGNs, however, suggests that this is not a common event.

We cross-matched the eight INOV AGNs with long-term variability studies in the COSMOS field to determine whether there is a relation between short-term and long-term variability. \citet{Decicco2015} used optical monitoring data from the VLT survey telescope to select AGN by variability. There are 5 months monitoring data covering the COSMOS field, which confirmed 67 AGNs with variability. However, their list of variable AGNs do not match with the eight variable AGNs found in this study. \citet{Simm2015,Simm2015} used 4-years of the Pan-STARRS data to examine AGN variability. Among our eight variable sources, three (095820.7+020213, 095834.0+024427, and 100151.1+020032) AGNs are matched with their samples. For these three AGNs, year-scale variability is found in several bands. So far, the result is mixed whether the INOV AGNs show long-term variability, and future investigation is necessary to reach to a firm conclusion.

The AGN long-term variability is attributed to the changes in accretion disk activities. A long-term variability of 0.1 mag over, say 30 days, can produce an INOV variability of about 0.003 mag. Detecting this kind of INOV variability can possibly be used to diagnose if an AGN is undergoing a long-term AGN variable activities. This kind of observation is beyond the scope of this study, but could be achieved with future observations with improved strategies and facilities.

There are two aspects where we can improve the study with the current instrument. One is to increase the exposure time to lower the photon noise, and another is to extend the observing period of the monitoring to catch intermittent AGN variation which was reported in other studies. We are now performing a longer term, higher S/N monitoring study and we hope to report the new  results in near future that are based on the improved observing strategy. A significant improvement in the INOV study sensitivity is also expected from future using a large, wide-field telescopes such as the Large Synoptic Survey Telescope (\citealt{Juric2015}), or high-cadence space missions such as the $Kepler$.

%%% ACKNOWLEDGMENTS (IF ANY) %%%%%%%%%%%%%%%%%%%%%%%%%%%%%%%%%%%%%%%%

\acknowledgments

We thank the anonymous referee for useful comments. This research made use of the KMTNet system operated by the Korea Astronomy and Space Science Institute (KASI) and data obtained at a host site of CTIO in Chile. This research was supported by Basic Science Research Program through the National Research Foundation of Korea (NRF) funded by the Ministry of Education (NRF-2017R1A6A3A04005158). We would like to thank CEOU (Center for the Exploration of the Origin of the Universe) members who gave helpful comments and support.

%%% CALL LIST OF REFERENCES (natbib STYLE) %%%%%%%%%%%%%%%%%%%%%%%%%%

\appendix

\section{List of observed AGNs\label{sec8}}

%%% TABLE %%%%%%%%%%%%%%%%%%%%%%%%%%%%%%%%%%%%%%%%%%%%%%%%%%%%%%%%%%%%%%%%%%%%%
\begin{table}[b!]
\renewcommand{\tabcolsep}{1.5mm}
\centering
\caption{List of observed AGNs}
\begin{tabular}{lcccccccllllllll}
\toprule
Name & X & M & R & S & R mag & z & $R_{i}$ &\multicolumn{8}{l}{1-p (C value)}\\
& & & & & & & &\multicolumn{4}{l}{2015.02.18} & \multicolumn{2}{l}{2015.03.21} & \multicolumn{2}{l}{2016.04.08} \\
& & & & & & & & B & V & R & I & B & R & B & R \\
\midrule
095621.66+024859.9&x&x&x&o& 19.1&0.95&-1&-99 (-99)&-99 (-99)&-99 (-99)&-99 (-99)&-99 (-99)&-99 (-99)&1.00 (7.23)&1.00 (11.7)\\
095642.33+020554.5&x&x&x&o& 19.4&1.03&-1&0.34 (1.04)&0.78 (1.40)&-99 (-99)&-99 (-99)&0.15 (1.02)&0.30 (0.97)&0.19 (1.19)&0.01 (0.73)\\
095642.71+024812.7&x&x&x&o& 19.7&2.00&-1&0.69 (1.53)&0.37 (1.17)&0.67 (1.40)&0.05 (0.66)&0.48 (1.12)&0.42 (0.79)&0.13 (1.12)&0.80 (1.53)\\
095657.63+011310.1&x&x&x&o& 20.2&2.33&-1&0.31 (1.00)&0.94 (1.90)&0.41 (1.16)&0.07 (0.78)&0.31 (1.06)&0.25 (1.11)&0.06 (1.00)&0.79 (1.55)\\
095659.32+015236.2&x&x&x&o& 20.6&1.55&-1&0.05 (0.62)&0.67 (1.40)&0.11 (0.74)&0.00 (0.37)&0.25 (1.23)&-99 (-99)&0.18 (1.24)&0.01 (0.74)\\
\bottomrule
\end{tabular}
\tabnote{X = X-ray selected AGN, M = MIR selected AGN, R = Radio selected AGN, S = SDSS DR7 quasar\\
The value "-1" in $R_{i}$ is source which position is out of the VLA field of view\\
The dof values used for the statistic are typically 4, 12, 10 for the 2015.02.18, 2016.03.21 and 2016.04.08 data respectively.}
\label{tab5}
\end{table}
\end{landscape}

\begin{landscape}
\begin{table}[b!]
\renewcommand{\tabcolsep}{1.5mm}
\centering
\begin{tabular}{lcccccccllllllll}
\toprule
Name & X & M & R & S & R mag & z & $R_{i}$ &\multicolumn{8}{l}{1-p (C value)}\\
& & & & & & & &\multicolumn{4}{l}{2015.02.18} & \multicolumn{2}{l}{2015.03.21} & \multicolumn{2}{l}{2016.04.08} \\
& & & & & & & & B & V & R & I & B & R & B & R \\
\midrule
095754.71+023832.9&x&o&x&o& 19.3&1.60&-99&0.30 (1.02)&0.26 (0.97)&0.37 (1.08)&0.63 (1.45)&0.13 (0.85)&0.02 (0.54)&0.74 (1.85)&0.61 (1.52)\\
095755.08+024806.6&x&o&x&o& 19.0&1.11&-99&0.11 (0.81)&0.01 (0.51)&0.78 (1.81)&0.30 (1.08)&0.32 (1.22)&0.05 (0.80)&0.51 (1.52)&0.06 (0.82)\\
095756.27+022344.8&x&x&o&x& 19.4&0.48&0.94&-99 (-99)&0.00 (0.41)&0.94 (2.33)&0.60 (1.46)&-99 (-99)&0.72 (1.23)&0.29 (1.29)&0.05 (0.85)\\
095756.52+022717.3&x&x&o&x& 21.7&0.73&2.21&-99 (-99)&-99 (-99)&-99 (-99)&0.00 (0.41)&-99 (-99)&-99 (-99)&-99 (-99)&-99 (-99)\\
095756.60+025530.1&x&o&x&x& 20.8&-99&-1&0.09 (0.72)&0.01 (0.41)&0.00 (0.17)&0.09 (0.74)&-99 (-99)&-99 (-99)&0.16 (1.15)&0.07 (0.89)\\
\bottomrule
\end{tabular}
\end{table}
\end{landscape}

\begin{landscape}
\begin{table}[b!]
\renewcommand{\tabcolsep}{1.5mm}
\centering
\begin{tabular}{lcccccccllllllll}
\toprule
Name & X & M & R & S & R mag & z & $R_{i}$ &\multicolumn{8}{l}{1-p (C value)}\\
& & & & & & & &\multicolumn{4}{l}{2015.02.18} & \multicolumn{2}{l}{2015.03.21} & \multicolumn{2}{l}{2016.04.08} \\
& & & & & & & & B & V & R & I & B & R & B & R \\
\midrule
095835.98+015157.1&o&o&o&o& 19.6&2.94&0.87&0.40 (1.06)&0.00 (0.20)&0.74 (1.56)&0.01 (0.50)&0.66 (1.10)&0.47 (1.25)&0.18 (1.24)&0.31 (1.15)\\
095837.36+023602.9&x&o&x&x& 21.0&1.20&-99&-99 (-99)&-99 (-99)&0.85 (1.68)&-99 (-99)&-99 (-99)&-99 (-99)&-99 (-99)&0.20 (1.01)\\
095839.26+020506.8&o&o&x&x& 20.1&1.22&-99&0.00 (0.28)&0.24 (0.87)&0.77 (1.43)&0.11 (0.82)&0.24 (0.98)&-99 (-99)&0.10 (1.07)&-99 (-99)\\
095840.02+021711.1&o&x&x&x& 21.2&1.02&-99&0.01 (0.47)&0.90 (2.10)&0.19 (0.85)&-99 (-99)&-99 (-99)&-99 (-99)&-99 (-99)&-99 (-99)\\
095841.80+015318.2&o&x&x&x& 20.1&1.78&-99&0.30 (1.08)&0.02 (0.53)&0.10 (0.82)&0.28 (0.99)&0.02 (0.87)&-99 (-99)&0.11 (1.09)&-99 (-99)\\
\bottomrule
\end{tabular}
\end{table}
\end{landscape}

\begin{landscape}
\begin{table}[b!]
\renewcommand{\tabcolsep}{1.5mm}
\centering
\begin{tabular}{lcccccccllllllll}
\toprule
Name & X & M & R & S & R mag & z & $R_{i}$ &\multicolumn{8}{l}{1-p (C value)}\\
& & & & & & & &\multicolumn{4}{l}{2015.02.18} & \multicolumn{2}{l}{2015.03.21} & \multicolumn{2}{l}{2016.04.08} \\
& & & & & & & & B & V & R & I & B & R & B & R \\
\midrule
095924.47+015954.4&o&o&x&o& 18.3&1.23&-99&-99 (-99)&0.52 (1.24)&-99 (-99)&-99 (-99)&0.11 (0.82)&0.26 (0.73)&0.00 (0.69)&-99 (-99)\\
095925.55+011532.7&x&x&x&o& 18.6&1.69&-1&0.56 (1.35)&0.59 (1.32)&0.57 (1.23)&0.10 (0.81)&0.57 (1.39)&0.62 (0.87)&0.08 (1.13)&0.17 (0.99)\\
095926.89+015341.5&o&o&x&x& 19.3&0.44&0.16&0.07 (0.64)&0.10 (0.73)&0.67 (1.29)&0.04 (0.63)&-99 (-99)&0.56 (1.43)&0.08 (1.16)&0.28 (1.20)\\
095928.33+021950.6&o&o&x&x& 20.6&1.47&-99&0.11 (0.77)&0.26 (0.91)&0.00 (0.17)&-99 (-99)&0.16 (0.92)&-99 (-99)&0.36 (1.35)&0.06 (0.83)\\
095928.45+015934.8&o&x&x&x& 21.3&1.16&-99&0.04 (0.53)&-99 (-99)&0.16 (0.90)&-99 (-99)&-99 (-99)&-99 (-99)&0.08 (1.12)&-99 (-99)\\
\bottomrule
\end{tabular}
\end{table}
\end{landscape}

\begin{landscape}
\begin{table}[b!]
\renewcommand{\tabcolsep}{1.5mm}
\centering
\begin{tabular}{lcccccccllllllll}
\toprule
Name & X & M & R & S & R mag & z & $R_{i}$ &\multicolumn{8}{l}{1-p (C value)}\\
& & & & & & & &\multicolumn{4}{l}{2015.02.18} & \multicolumn{2}{l}{2015.03.21} & \multicolumn{2}{l}{2016.04.08} \\
& & & & & & & & B & V & R & I & B & R & B & R \\
\midrule
095959.50+013151.7&x&x&o&x& 21.0&1.01&0.93&0.04 (0.55)&0.25 (1.03)&0.09 (0.65)&0.02 (0.56)&-99 (-99)&-99 (-99)&0.09 (0.97)&0.37 (1.26)\\
095959.63+024608.8&x&x&o&x& 17.0&0.16&0.64&0.53 (1.38)&0.98 (2.66)&-99 (-99)&-99 (-99)&0.83 (1.44)&0.18 (0.96)&0.97 (2.37)&0.57 (2.13)\\
100001.45+024844.7&x&o&x&o& 19.4&0.76&0.93&0.81 (1.68)&0.39 (1.08)&0.92 (1.68)&0.64 (1.47)&0.13 (1.00)&0.60 (1.08)&0.24 (1.32)&0.75 (1.51)\\
100001.76+024016.1&x&o&x&x& 21.4&0.01&-99&-99 (-99)&-99 (-99)&0.25 (0.82)&-99 (-99)&-99 (-99)&-99 (-99)&0.02 (0.88)&0.09 (0.97)\\
100005.44+023029.2&x&x&o&x& 20.6&0.74&1.97&-99 (-99)&0.01 (0.41)&0.19 (0.95)&0.07 (0.72)&-99 (-99)&-99 (-99)&-99 (-99)&0.58 (1.56)\\
\bottomrule
\end{tabular}
\end{table}
\end{landscape}

\begin{landscape}
\begin{table}[b!]
\renewcommand{\tabcolsep}{1.5mm}
\centering
\begin{tabular}{lcccccccllllllll}
\toprule
Name & X & M & R & S & R mag & z & $R_{i}$ &\multicolumn{8}{l}{1-p (C value)}\\
& & & & & & & &\multicolumn{4}{l}{2015.02.18} & \multicolumn{2}{l}{2015.03.21} & \multicolumn{2}{l}{2016.04.08} \\
& & & & & & & & B & V & R & I & B & R & B & R \\
\midrule
100050.43+021842.0&x&x&o&x& 21.3&0.71&1.09&-99 (-99)&-99 (-99)&0.10 (0.71)&0.00 (0.38)&-99 (-99)&-99 (-99)&-99 (-99)&0.14 (1.00)\\
100051.92+015919.5&o&o&x&x& 21.6&2.24&-99&0.11 (0.70)&0.18 (0.83)&0.30 (1.05)&-99 (-99)&-99 (-99)&-99 (-99)&0.18 (1.25)&-99 (-99)\\
100054.50+021605.1&o&x&x&x& 19.4&-99&-99&-99 (-99)&-99 (-99)&0.32 (1.07)&0.12 (0.85)&-99 (-99)&-99 (-99)&-99 (-99)&-99 (-99)\\
100054.79+014602.5&x&o&x&x& 20.5&0.37&0.68&-99 (-99)&0.47 (1.32)&0.03 (0.61)&0.00 (0.38)&-99 (-99)&-99 (-99)&-99 (-99)&0.57 (1.38)\\
100055.39+023441.4&o&o&x&o& 19.6&1.40&0.24&-99 (-99)&-99 (-99)&-99 (-99)&0.04 (0.63)&0.05 (1.06)&0.05 (0.87)&0.02 (0.92)&0.10 (0.92)\\
\bottomrule
\end{tabular}
\end{table}
\end{landscape}

\begin{landscape}
\begin{table}[b!]
\renewcommand{\tabcolsep}{1.5mm}
\centering
\begin{tabular}{lcccccccllllllll}
\toprule
Name & X & M & R & S & R mag & z & $R_{i}$ &\multicolumn{8}{l}{1-p (C value)}\\
& & & & & & & &\multicolumn{4}{l}{2015.02.18} & \multicolumn{2}{l}{2015.03.21} & \multicolumn{2}{l}{2016.04.08} \\
& & & & & & & & B & V & R & I & B & R & B & R \\
\midrule
100120.26+023341.4&o&o&x&o& 19.9&1.83&-99&0.18 (0.92)&0.81 (1.60)&0.14 (0.78)&0.03 (0.64)&0.00 (0.56)&0.20 (0.99)&0.00 (0.67)&0.41 (1.34)\\
100122.46+020112.5&x&x&o&x& 20.1&0.42&2.52&-99 (-99)&-99 (-99)&-99 (-99)&0.74 (1.62)&-99 (-99)&-99 (-99)&-99 (-99)&0.33 (1.09)\\
100124.10+024936.4&x&x&o&x& 21.7&0.82&2.52&-99 (-99)&-99 (-99)&-99 (-99)&0.19 (0.96)&-99 (-99)&-99 (-99)&-99 (-99)&-99 (-99)\\
100124.74+015738.8&o&o&x&x& 21.2&1.17&-99&-99 (-99)&0.33 (0.98)&0.05 (0.61)&-99 (-99)&-99 (-99)&-99 (-99)&0.31 (1.50)&0.06 (0.89)\\
100124.86+022031.9&o&x&x&x& 20.5&1.70&-99&-99 (-99)&-99 (-99)&-99 (-99)&-99 (-99)&0.01 (0.67)&-99 (-99)&0.35 (1.37)&-99 (-99)\\
\bottomrule
\end{tabular}
\end{table}
\end{landscape}

\begin{landscape}
\begin{table}[b!]
\renewcommand{\tabcolsep}{1.5mm}
\centering
\begin{tabular}{lcccccccllllllll}
\toprule
Name & X & M & R & S & R mag & z & $R_{i}$ &\multicolumn{8}{l}{1-p (C value)}\\
& & & & & & & &\multicolumn{4}{l}{2015.02.18} & \multicolumn{2}{l}{2015.03.21} & \multicolumn{2}{l}{2016.04.08} \\
& & & & & & & & B & V & R & I & B & R & B & R \\
\midrule
100159.43+023935.7&x&o&x&x& 19.9&1.37&-99&0.37 (1.13)&0.92 (1.79)&0.98 (2.33)&0.02 (0.52)&0.02 (0.80)&0.56 (1.50)&0.61 (1.57)&0.03 (0.82)\\
100159.79+022641.7&o&o&x&x& 19.2&2.03&-99&-99 (-99)&0.24 (0.94)&0.53 (1.18)&0.05 (0.63)&0.53 (1.22)&-99 (-99)&-99 (-99)&-99 (-99)\\
100159.86+013135.3&x&o&x&x& 21.4&0.30&-99&0.12 (0.76)&0.13 (0.76)&-99 (-99)&-99 (-99)&-99 (-99)&-99 (-99)&-99 (-99)&0.11 (1.00)\\
100201.52+020329.4&o&o&x&o& 18.2&2.02&-99&0.73 (1.66)&0.97 (2.17)&-99 (-99)&-99 (-99)&0.80 (1.65)&0.58 (1.05)&0.21 (1.16)&-99 (-99)\\
100202.23+024157.8&x&o&x&x& 21.2&1.53&-99&0.14 (0.78)&-99 (-99)&0.36 (0.99)&-99 (-99)&-99 (-99)&-99 (-99)&0.39 (1.38)&-99 (-99)\\
\bottomrule
\end{tabular}
\end{table}
\end{landscape}

\begin{landscape}
\begin{table}[b!]
\renewcommand{\tabcolsep}{1.5mm}
\centering
\begin{tabular}{lcccccccllllllll}
\toprule
Name & X & M & R & S & R mag & z & $R_{i}$ &\multicolumn{8}{l}{1-p (C value)}\\
& & & & & & & &\multicolumn{4}{l}{2015.02.18} & \multicolumn{2}{l}{2015.03.21} & \multicolumn{2}{l}{2016.04.08} \\
& & & & & & & & B & V & R & I & B & R & B & R \\
\midrule
100246.93+015301.2&x&x&o&x& 19.5&0.52&0.54&-99 (-99)&0.26 (0.90)&-99 (-99)&-99 (-99)&-99 (-99)&-99 (-99)&0.32 (1.37)&0.59 (1.51)\\
100247.00+024538.7&x&x&o&x& 20.8&0.47&1.10&-99 (-99)&-99 (-99)&0.48 (1.26)&0.89 (1.88)&-99 (-99)&-99 (-99)&-99 (-99)&-99 (-99)\\
100249.00+021750.7&x&o&x&x& 21.5&2.50&-99&-99 (-99)&0.92 (1.46)&-99 (-99)&-99 (-99)&-99 (-99)&-99 (-99)&-99 (-99)&0.21 (1.08)\\
100249.32+023746.6&x&o&x&x& 19.9&0.01&2.24&0.50 (1.21)&-99 (-99)&-99 (-99)&-99 (-99)&0.04 (0.77)&0.37 (1.36)&0.06 (1.00)&-99 (-99)\\
100249.92+021732.4&x&o&x&x& 21.0&1.60&-99&0.24 (0.87)&0.89 (1.72)&0.65 (1.12)&-99 (-99)&-99 (-99)&-99 (-99)&-99 (-99)&0.61 (1.34)\\
\bottomrule
\end{tabular}
\end{table}
\end{landscape}

\begin{landscape}
\begin{table}[h!]
\renewcommand{\tabcolsep}{1.5mm}
\centering
\begin{tabular}{lcccccccllllllll}
\toprule
Name & X & M & R & S & R mag & z & $R_{i}$ &\multicolumn{8}{l}{1-p (C value)}\\
& & & & & & & &\multicolumn{4}{l}{2015.02.18} & \multicolumn{2}{l}{2015.03.21} & \multicolumn{2}{l}{2016.04.08} \\
& & & & & & & & B & V & R & I & B & R & B & R \\
\midrule
100358.73+014513.6&x&x&x&o& 20.1&1.41&-1&0.17 (0.86)&0.02 (0.52)&0.44 (1.07)&0.16 (0.89)&0.26 (1.15)&-99 (-99)&0.63 (1.54)&0.00 (0.61)\\
100407.21+012039.0&x&x&x&o& 20.6&1.84&-1&0.13 (0.82)&0.71 (1.37)&0.83 (1.39)&0.19 (0.96)&0.97 (1.96)&-99 (-99)&-99 (-99)&0.93 (1.68)\\
100409.29+024946.4&x&x&x&o& 19.1&1.59&-1&-99 (-99)&-99 (-99)&-99 (-99)&-99 (-99)&-99 (-99)&0.32 (0.99)&0.00 (0.81)&0.50 (1.30)\\
100421.02+013647.4&x&x&x&o& 18.5&1.03&-1&0.25 (1.04)&0.75 (1.49)&0.99 (2.72)&0.00 (0.39)&0.91 (2.05)&0.20 (1.05)&0.52 (1.55)&0.31 (1.19)\\
100423.49+030658.9&x&x&x&o& 19.0&0.91&-1&0.05 (0.64)&0.08 (0.68)&0.02 (0.50)&0.19 (1.01)&0.57 (1.13)&0.49 (0.99)&0.33 (1.37)&0.31 (1.16)\\
100429.00+025209.2&x&x&x&o& 18.5&0.87&-1&0.07 (0.69)&0.54 (1.29)&-99 (-99)&0.00 (0.48)&0.07 (0.58)&0.31 (1.24)&0.20 (1.17)&0.50 (1.40)\\
100434.37+025011.0&x&x&x&o& 18.4&1.21&-1&-99 (-99)&1.00 (6.45)&1.00 (7.94)&-99 (-99)&1.00 (4.99)&-99 (-99)&1.00 (7.39)&1.00 (10.0)\\
\bottomrule
\end{tabular}
\end{table}
\end{landscape}
%%%%%%%%%%%%%%%%%%%%%%%%%%%%%%%%%%%%%%%%%%%%%%%%%%%%%%%%%%%%%%%%%%%%%%%%%%%%%%%
%\end{landscape}

\end{document}